\documentclass[draft]{agujournal}
\usepackage[T1]{fontenc}
\usepackage[latin9]{inputenc}
\usepackage{color}
\usepackage{array}
\usepackage{float}
\usepackage{textcomp}
\usepackage{url}
\usepackage{multirow}
\usepackage{amstext}
\usepackage{graphicx}

\makeatletter

\providecommand{\tabularnewline}{\\}

\journalname{WRR}
\usepackage{lmodern}
\usepackage[T1]{fontenc}
\usepackage{bm}

\makeatother

\begin{document}

\title{Combining Physically-Based Modeling and Deep Learning for Fusing
GRACE Satellite Data: Can We Learn from Mismatch?}

\authors{\authors{Alexander Y. Sun\affil{1},
Bridget R. Scanlon\affil{1},
Zizhan Zhang\affil{2},
David Walling\affil{3}, Soumendra N. Bhanja\affil{4},
Abhijit Mukherjee\affil{4},
Zhi Zhong\affil{1}
}}

\affiliation{1}{Bureau of Economic Geology, Jackson School of Geosciences, The University of Texas at Austin, Austin, TX, USA}

\affiliation{2}{State Key Laboratory of Geodesy and Earth's Dynamics, Institute of Geodesy and Geophysics, Chinese Academy of Sciences, Wuhan 430077, China}
\affiliation{3}{Texas Advanced Computing Center, The University of Texas at Austin, Austin, TX, USA}
\affiliation{4}{Department of Geology and Geophysics, Indian Institute of Technology Kharagpur, West Bengal 721302, India}

\correspondingauthor{A. Y. Sun}{alex.sun@beg.utexas.edu}
\begin{abstract}
Global hydrological and land surface models are increasingly used
for tracking terrestrial total water storage (TWS) dynamics, but the
utility of existing models is hampered by conceptual and/or data uncertainties
related to various underrepresented and unrepresented processes, such
as groundwater storage. The gravity recovery and climate experiment
(GRACE) satellite mission provided a valuable independent data source
for tracking TWS at regional and continental scales. Strong interests
exist in fusing GRACE data into global hydrological models to improve
their predictive performance. Here we develop and apply deep convolutional
neural network (CNN) models to learn the spatiotemporal patterns of
mismatch between TWS anomalies (TWSA) derived from GRACE and those
simulated by NOAH, a widely used land surface model. Once trained,
our CNN models can be used to correct the NOAH simulated TWSA without
requiring GRACE data, potentially filling the data gap between GRACE
and its follow-on mission, GRACE-FO. Our methodology is demonstrated
over India, which has experienced significant groundwater depletion
in recent decades that is nevertheless not being captured by the NOAH
model. Results show that the CNN models significantly improve the
match with GRACE TWSA, achieving a country-average correlation coefficient
of 0.94 and Nash-Sutcliff efficient of 0.87, or 14\% and 52\% improvement
respectively over the original NOAH TWSA. At the local scale, the
learned mismatch pattern correlates well with the observed in situ
groundwater storage anomaly data for most parts of India, suggesting
that deep learning models effectively compensate for the missing groundwater
component in NOAH for this study region. 
\end{abstract}

\section{Introduction}

Terrestrial total water storage (TWS) is a key element of the global
hydrological cycle, affecting both water and energy budgets \citep{Rodell2001}.
Tracking the TWS on a periodic basis was historically difficult because
of the lack of reliable in situ observations \citep{seneviratne2004inferring},
a situation that is still true in most countries. The gravity recovery
and climate experiment (GRACE) satellite mission provided unprecedented
tracking of the global TWS dynamics during its 15-year mission (2002-2017).
GRACE enabled remote sensing of TWS anomalies (TWSA) (i.e., variations
from a long-term mean) at regional to continental scales (> 100,000
km\textsuperscript{2}). The availability of such information has
had a profound impact on the development and validation of regional
and global hydrological models, which are increasingly being used
to assess changes in the hydrological cycle under current and future
climate conditions. These physically-based, semi-distributed hydrological
models are built on mathematical abstractions of physical processes
that govern the movement and storage of water, as well as land surface
energy partitioning in certain models, in space and time. Despite
its coarse resolution, GRACE provides a \textquotedblleft big picture\textquotedblright{}
check of model simulated TWS variations and thus represents a valuable
independent source of information for diagnosing and improving the
model performance. So far, GRACE data has been used in model calibration
and parameter estimation \citep{werth2010calibration,lo2010,Milzow2011,Sun2012}
and data assimilation \citep{Houborg2012,Li2012,Dijk2014,Girotto2016,schumacher2016systematic,Khaki2017}.
While results of these studies all indicate that the assimilation
of GRACE data generally improves model skills, the improvements may
be limited by uncertainties in model parameters and structures (e.g.,
missing deep groundwater storage and agricultural irrigation), as
well as assumptions underlying data assimilation schemes (e.g., a
priori specified spatial and temporal error covariance structures)
\citep{Girotto2016}. Calibration against an imperfect model structure
using inaccurate error models may lead to information loss and greater
propensity for forecast error \citep{gupta2014debates}. A recent
study compared TWSA trends obtained from seven global hydrological
models with those derived from GRACE over 186 global river basins
\citep{Scanlon2018}. Their results indicate a large spread in model
results and poor correlation between models and GRACE, which were
attributed by the authors to the lack of surface water and groundwater
storage components in most land surface models (LSMs), low storage
capacity in all models, uncertainties in climate forcing, and lack
of representation of human intervention in most LSMs. 

Unlike physically-based models, pure data-driven methods (black box
models) seek to establish a regression model between climate forcing
(e.g., precipitation and temperature) and GRACE TWS \citep{Long2014,Humphrey2017,Seyoum2017},
or between TWS and its various components \citep{Sun2013,zhang2016grace,Miro2018}.
Data-driven models are suitable for applications where there are plenty
of observations but a complete understanding of the underlying physical
processes is lacking. A common criticism of black box models, however,
is related to their lack of interpretability and generalizability\textemdash a
regression model trained on the premise of a strong correlation between
predictors and the predictand may give unreliable results whenever
and wherever such correlation is weak. In addition, pure data-driven
models often do not integrate the full stack of information (e.g.,
soil property, topography, and vegetation types) that is normally
represented in physically-based models and therefore are only limited
to simulating certain aspects (e.g., interannual variations) of a
physical process. It is thus desirable to apply knowledge gained from
decades of physical-based modeling to inform the development of data-driven
models. These hybrid physical science and data science methods will
help to bridge and thus benefit hypothesis-driven and data-driven
discoveries \citep{karpatne2017theory}.

In this work we apply a hybrid approach that combines the strengths
of physically-based modeling and deep learning. Specifically, we use
deep convolutional neural networks (CNN), which are a special class
of artificial neural networks, to learn the spatiotemporal patterns
of \textquotedblleft mismatch\textquotedblright{} between the TWSA
simulated by an LSM and that observed by GRACE. Here the term mismatch
broadly refers to the difference either between two datasets or between
model simulations and observations. The learned mismatch patterns
are then fed back to the LSM to compensate for deficiencies in the
LSM. That means once trained and validated, the CNN model may be used
to predict the observed TWSA without requiring GRACE TWSA as inputs,
thus potentially filling the data gap between GRACE and its follow-on
mission (GRACE-FO). In the same fashion, the trained CNN model may
also be used to reconstruct TWSA for the pre-GRACE era. The basic
principle underlying our hybrid modeling approach is similar to that
behind data assimilation methods, both exploiting mismatch patterns
between predicted and observed variables. However, the assimilation
part of our hybrid method is driven by deep learning models that set
the current state-of-the-art in computer vision, and not limited by
the Gaussian-like unimodal error distribution commonly assumed in
many data assimilation schemes. On the other hand, the spatiotemporal
propagation part of our method is driven by a physically based LSM,
mitigating the lack of spatial continuity and physical interpretation
in purely data-driven statistical models. 

As a case study, we demonstrate our hybrid approach over India, where
irrigation-induced groundwater depletion has been confirmed by GRACE
and in situ studies \citep{Rodell2004,Chen2014,Long2016,macdonald2016groundwater},
but is not well resolved in many contemporary LSMs. We evaluate the
performance of three different types of CNN models, driven under different
predictor combinations. Compared to the original LSM, we show that
all CNN models considered here significantly improve the performance
of the corrected LSM model, both at the country and grid scales. In
the following, Section 2 describes the study area and data used, Section
3 provides details on the technical approach, and results and discussions
are given in Section 4. 

\section{Data and Data Processing}

\subsection{Description of the study area, India}

A large part of the annual rainfall budget over the Indian subcontinent
can be attributed to the Indian Summer monsoon (ISM), which results
from interactions of several complex atmospheric processes evolving
over many different spatiotemporal scales and is modulated by the
steep topography of the Himalayas \citep{Bookhagen2010}. The entire
Indian region (except for the southern part) receives maximum precipitation
during the monsoon season, which typically lasts from June to September.
At the country level, the average rainfall received during the monsoon
season is 85 cm, amounting to about 78\% of the annual rainfall \citep{mooley1984fluctuations}.
In the southern part of the country, the monsoon season extends to
October, sometimes even to November \citep{Bhanja2016}. 

India depends heavily on groundwater resources. Groundwater storage
is a function of climatic variables such as precipitation and evaporation,
particularly in areas with shallow groundwater tables \citep{Bhanja2016}.
The Indus\textendash Ganges\textendash Brahmaputra systems, which
together drain the northern Indian plains, form a regional alluvial
aquifer system that is regarded as one of the most productive aquifers
of the world; on the other hand, groundwater is available in a limited
extent within the weathered zone and underlying fractured aquifers
within the remaining two-thirds of the country \citep{mukherjee2015groundwater}.
Irrigation withdrawal accounts for over 90\% of the total groundwater
uses \citep{IndiaBoard2014}. Overuse of groundwater beyond its potential
has caused pronounced groundwater depletion in northwest India, including
the states of Punjab, Haryana and Delhi, and Rajasthan (Figure \ref{fig:1}).
The country has established a dense in situ groundwater monitoring
network. Groundwater level measurements are taken on a seasonal basis
in January, April/May, August, and November, from a network of piezometers
(4,939) and non-pumping observation wells (10,714) that are typically
screened in the first available aquifer below ground surface \citep{Bhanja2016}. 
\begin{center}
\begin{figure}[h]
\centering{}\includegraphics[width=4in]{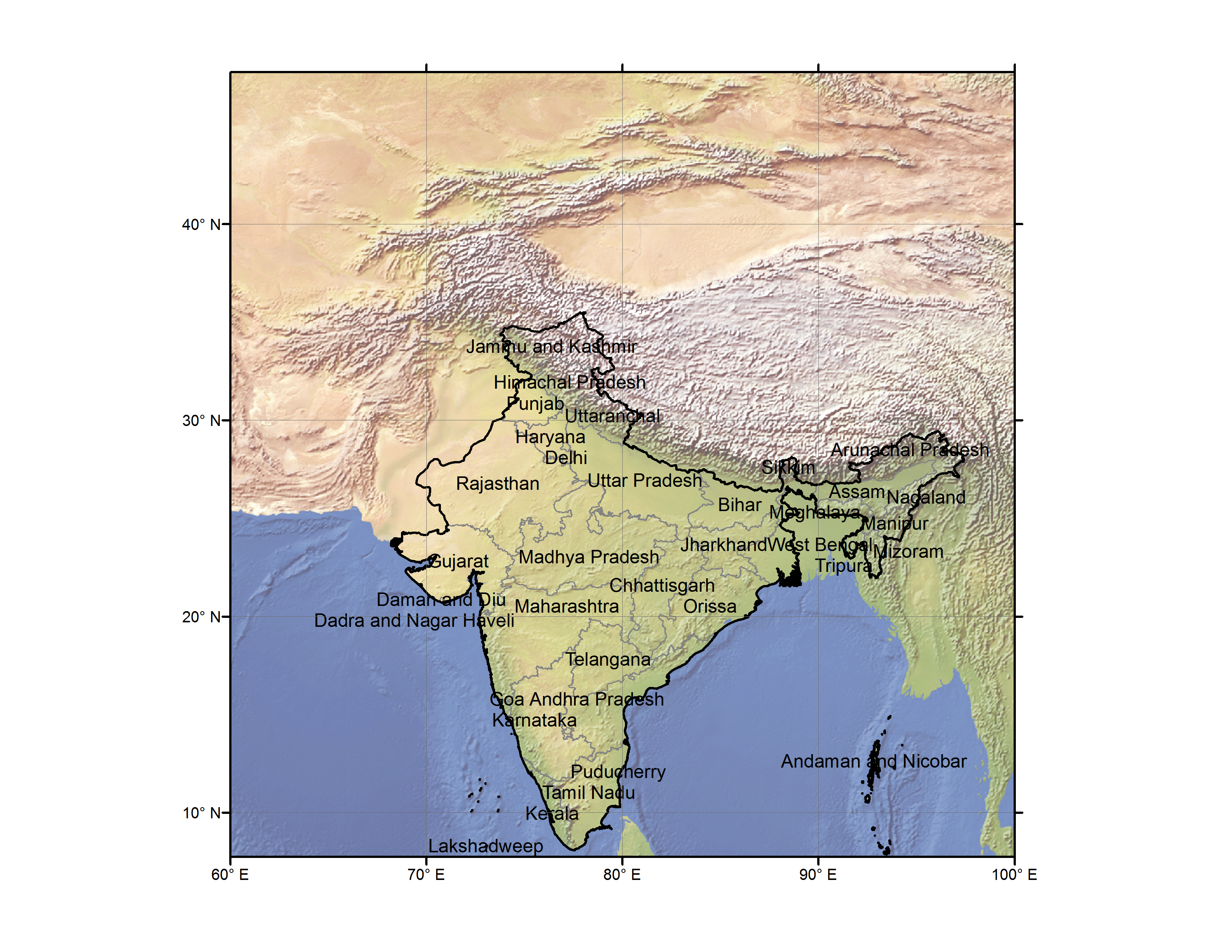}\caption{Map of study area (latitude: 7.75\textendash 47.75$\text{\textdegree}$, longitude:
60\textendash 100$\text{\textdegree}$), where India is bounded by the dark
solid line. During training, data corresponding to the entire square
area is used to reduce potential boundary effects and increase information
content for training. \label{fig:1}}
\end{figure}
\par\end{center}

The extensive in situ groundwater monitoring coverage shall provide
additional information for cross-validating patterns learned by the
deep CNN models. This study uses the in situ groundwater dataset published
recently by \citet{Bhanja2016}, which consists of 3,989 wells that
were selected to have temporal continuity (i.e., at least 3 out of
4 seasonal data should be available in all years). The authors derived
groundwater storage anomalies from water level measurements by using
specific yield values corresponding to 12 major river basins in the
country. The temporal coverage of the dataset is from January 2005
to November 2014. More details on the data processing and quality
control can be found in \citet{Bhanja2016}. 

Besides groundwater, the impact of surface water is relatively high
along Indus River and Ganges River, but is generally small in the
area of severe groundwater depletion in northwest India \citep{getirana2017rivers}.

\subsection{GRACE-derived TWSA}

This study uses the monthly mascon TWSA product (RL-05) released by
Jet Propulsion Laboratory (JPL) (\url{https://grace.jpl.nasa.gov}),
which has a 0.5$\text{\textdegree}$\texttimes 0.5$\text{\textdegree}$ grid resolution,
but inherently represents 3$\text{\textdegree}$\texttimes 3$\text{\textdegree}$ equal-area
caps \citep{Watkins2015}. The period of study covers from April 2002
to December 2016. Uncertainty in GRACE data is related to both measurement
and leakage errors, leading to potential signal loss \citep{Wiese2016}.
Measurement errors are related to, for example, system-noise error
in the inter-satellite range-rate and accelerometer error \citep{swenson2003estimated}.
Leakage errors arise because boundaries of hydrological basins generally
do not conform to the boundaries of the mascon elements and because
leakage across land/ocean boundaries (i.e., from mascons that cover
both land and ocean). For this work, we applied the gain factor (scaling
factor) distributed with the JPL mascon to compensate for the signal
loss. The gain factor, when combined with coastal line resolution
improvement, was shown to reduce leakage errors associated with mass
balance of large river basins (>160,000 km\textsuperscript{2}) by
an amount of 0.6\textendash 1.5 mm equivalent water height averaged
globally \citep{Wiese2016}. We obtained the total uncertainty bound
of monthly TWSA for the study region by combining the measurement
error released by JPL with the estimated leakage error. The leakage
error was estimated using the method of \citet{Wiese2016}. 

\subsection{NOAH land surface model}

The NOAH LSM from NASA\textquoteright s global land data assimilation
system (GLDAS) \citep{Rodell2004} has been extensively used in previous
GRACE studies. Like many other LSMs, NOAH maintains surface energy
and water balances and simulates the exchange of water and energy
fluxes at soil-atmosphere interface \citep{ek2003implementation}.
NOAH does not simulate surface water storage (SWS) (e.g., in rivers,
lakes, and wetlands) and surface runoff routing, nor does it account
for deep groundwater storage and human intervention. The roles of
SWS and GWS can be significant in various parts of the study area,
as mentioned previously. For this study, the monthly forcing (total
precipitation and average air temperature at 2m) and outputs of NOAH
V2.1 (0.25$\text{\textdegree}$\texttimes 0.25$\text{\textdegree}$) were downloaded from
NASA's EarthData site (\url{http://earthdata.nasa.gov}). The NOAH-simulated
TWS was calculated by summing soil moisture in all four soil layers
(spanning from 0\textendash 200 cm depth), accumulative snow water,
and total canopy water storage (the contribution of canopy water is
typically negligible but is included for completeness). To be consistent
with the GRACE TWSA processing, the long-term mean from January 2004
to December 2009 was subtracted from NOAH TWS to obtain the simulated
TWSA.

\section{Methodology}

\subsection{Model and GRACE TWSA mismatch}

TWS is the sum of the following components \citep{Scanlon2018}:
\begin{equation}
\text{TWS}=\text{SnWS}+\text{CWS}+\text{SWS}+\text{SMS}+\text{GWS},
\end{equation}
where SnWS represents snow water storage, CWS is canopy water storage,
SWS is surface water storage, SMS is soil moisture storage, and GWS
is groundwater storage. We define the difference or mismatch between
NOAH-simulated TWSA and GRACE TWSA at time $t$ as
\begin{equation}
S(t)=\text{TWSA}{}_{\text{NOAH}}(t)-\text{TWSA}{}_{\text{GRACE}}(t),\label{eq:2}
\end{equation}
where the mismatch $S(t)$, which varies in both space and time, may
be related to two types of errors, (a) systematic error or bias caused
by either missing processes or uncertain conceptualization in NOAH
(e.g., omission of GWS), and (b) random error related to uncertain
data and model parameters. For the purpose of this work, we use CNN
models to learn a functional relationship between $S(t)$ and its
predictors $X$ by solving a regression problem
\begin{equation}
f:\:X\rightarrow S,
\end{equation}
where $f=f(X,{\bf w})$ is a CNN model; ${\bf w}$ denotes the network
parameters to be solved by using $\{X_{i},S_{i}\}_{i=1}^{N}$ as training
data, where $i=1\dots N$ is the index of training samples, $X_{i}=\{x^{j}\}_{j=1}^{M}$
is a set of input samples from $M$ different predictors $x_{j}\:(j=1,\dots M)$,
and $S_{i}$ are samples of $S(t)$ obtained by using Eq. \ref{eq:2}.
After training and validation, the CNN model can be used to predict
and, thus, give corrected TWSA without requiring GRACE data. 

\begin{figure}
\noindent \centering{}\includegraphics[height=5in]{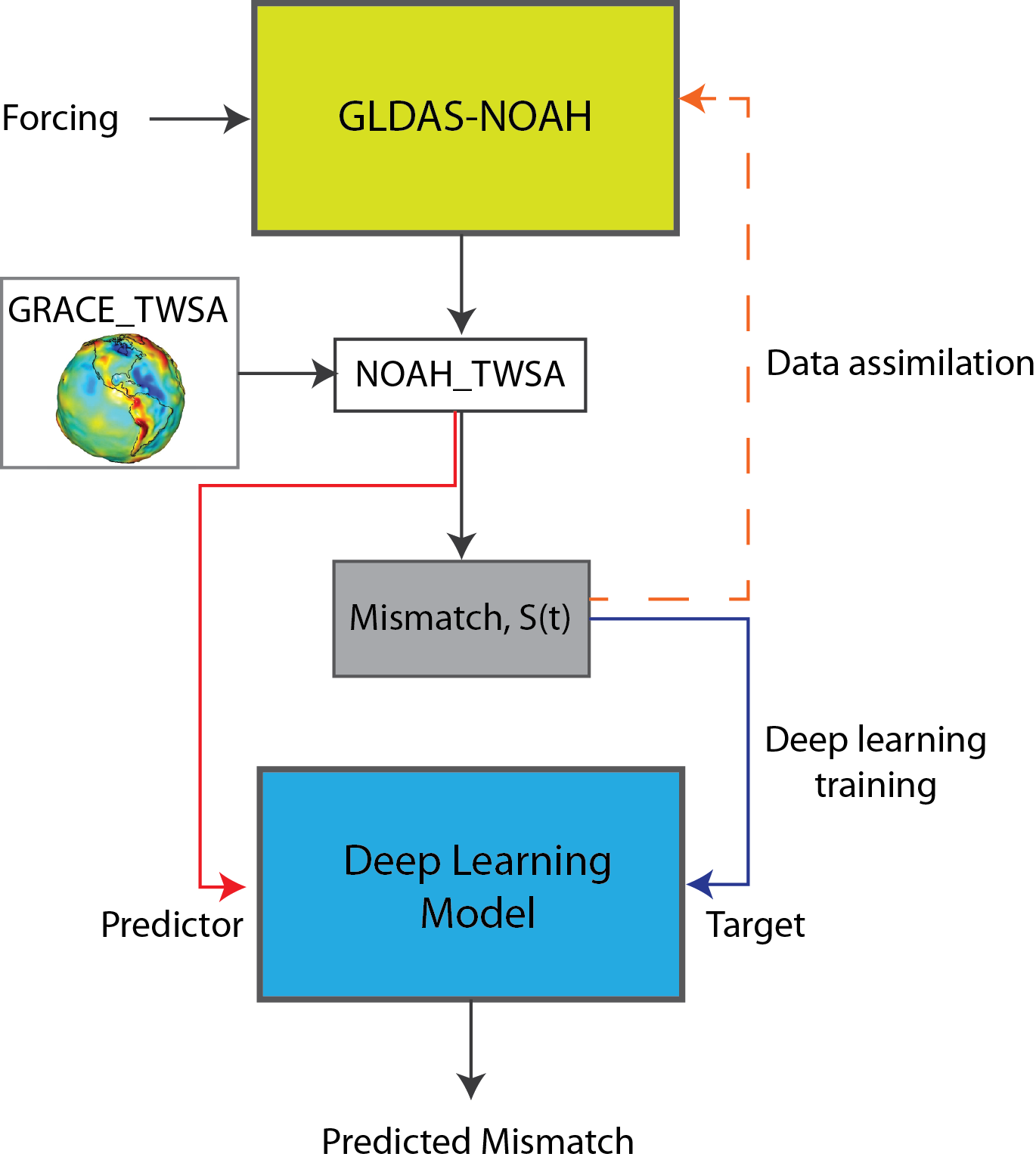}\caption{Illustration of the flow of information from GLDAS-NOAH and GRACE
to the deep learning model. Here the observed mismatch $S(t)$ (blue
solid line) is only used to train the CNN deep learning model and
is no longer required after the model is trained. NOAH TWSA is the
base predictor (red solid line). Other predictors may include precipitation
and temperature. The dashed arrow line indicates that the same $S(t)$
is also used for GRACE data assimilation studies. \label{fig:2}}
\end{figure}

Figure \ref{fig:2} further illustrates the relations among NOAH,
GRACE, and the deep learning model, and the proposed workflow. The
deep learning workflow (solid line) is similar to that used in the
traditional data assimilation (dashed line), both exploiting the residual
between model and observations. The main difference is that in deep
learning the GRACE TWSA data is not used to correct the model states
but to train a regression model for predicting the mismatch, circumventing
challenges related to calibrating a conceptually uncertain physical
model. Details on the design and architecture of the CNN models are
provided in the subsection below. 

\subsection{Design and architectures of CNN deep learning models}

CNN, originally introduced by LeCun \citep{LeCun1989,LeCun1995},
is symbolic of the modern deep learning era that began around 2006
\citep{schmidhuber2015deep}. CNNs and their variants have been extensively
used in image classification and are behind several high-profile deep
learning model architectures that have won the ImageNet Large Scale
Visual Recognition Challenge (ILSVRC) in recent years \citep{Simonyan2014,LeCun2015,szegedy2015,he2016deep}.
The design of CNN was inspired by the human visual cortex, aiming
to extract subtle features embedded in the inputs. As its name suggests,
CNN applies discrete convolution operations to project an input image
(or a stack of images) onto a hierarchy of feature maps, which may
be thought of as nonlinear transformations of the input. In practice,
a CNN deep learning model architecture includes the input, output,
and a series of hidden convolution layers in between to extract spatial
features (e.g., edges and corners) from each layer's input. Thus,
by design CNN models are highly suitable for learning multiscale spatial
patterns from multisource gridded data, which is a challenging problem
to solve using the traditional multilayer perception neural network
models that do not scale well on images. 

In a convolution operation, a moving window, commonly referred to
as a filter or kernel, is used to scan along each dimension of the
input image, with possible strides between the moves (a stride defines
the number of rows/columns to skip). For each move, a dot product
is taken between the filter parameters and the underlying input image
patch, leading to a feature map at the end of scanning. The dimensions
(width $W$ and height $H$) of a feature map are related to its input
as
\begin{equation}
W=(W_{in}-D_{F}+2D_{P})/D_{S}+1,\quad H=(H_{in}-D_{F}+2D_{P})/D_{S}+1,\label{eq:4}
\end{equation}

\noindent where $W_{in}$ and $H_{in}$ are dimensions of the input
image, $W$ and $H$ are dimensions of the feature map, $D_{F}$ is
the filter dimension, $D_{S}$ is the stride size, and $D_{P}$ is
padding size. Filter dimensions and stride sizes are commonly kept
the same for both dimensions. Eq. \ref{eq:4} suggests that the dimensions
of a feature map become progressively smaller after each convolution
operation. Zero-padding may be used to add zeros around the edges
of the output feature map (i.e., $D_{P}$ in Eq. \ref{eq:4}) to preserve
the input dimensions. 

CNN naturally achieves sparsity because each pixel in a feature map
only connects to a small region in its input layer. Also, by applying
the same filter to scan the entire input image, the filter parameters
are shared and the resulting feature map is equivariant to shifts
in inputs. Specifically, the units of a convolutional layer $l$,
$A_{j}^{(l)}$, is related to feature maps of its preceding layer
$l-1$, $A_{i}^{(l-1)}\:(i=1,\dots,M^{(l-1)})$, by \citep{goodfellow2016deep}

\begin{equation}
A_{j}^{(l)}=g\left(\sum_{i=1}^{M^{(l-1)}}A_{i}^{(l-1)}\oplus k_{ij}^{(l)}+b_{j}^{(l)}\right),\label{eq:5}
\end{equation}
where $M^{(l-1)}$ is the number of feature maps in layer $l-1$,
$\oplus$ denotes the convolution operator, $k_{ij}^{(l)}$ are the
filter parameters, $b_{j}^{(i)}$ are the bias parameters, and $g(\cdot)$
is the activation function. Eq. (\ref{eq:5}) shows that CNN involves
a hierarchy of feature maps, with each layer learning from its preceding
layer. When $l=1$ (i.e., the first hidden layer), its input layer
simply becomes the actual input image(s). The Rectified Linear Unit
(ReLU) function 
\begin{equation}
g(x)=\max(0,x)
\end{equation}

\noindent is commonly used as the activation function for hidden CNN
layers, which is less costly to compute than other nonlinear functions
(e.g., sigmoid) and is shown to improve the CNN training speed significantly
\citep{goodfellow2016deep}. In regression problems, the linear function
or hyperbolic tangent function (\text{tanh}) are often used as the
activation functions for the output layer to generate solution in
the real domain. The total number of CNN parameters (weights and biases)
is determined by the number of filters, filter dimensions, and stride
dimensions, which are considered hyperparameters of the CNN model
design and may be tuned during training. 

In addition to convolution operation, other commonly used CNN layer
operations include pooling, dropout, and batch normalization. Pooling
aggregates information in each moving window to further reduce the
size of feature maps. For example, max pooling selects the maximum
element in a pooling window. Dropout operation randomly leaves out
certain number of hidden neurons during training so that the net effect
is to prevent the network from overfitting; thus, it is regarded as
a regularization technique. Batch normalization performs normalization
on hidden layers to improve network training speed and stability \citep{goodfellow2016deep}. 

Figure \ref{fig:3} shows a high-level, architectural diagram of CNN
deep learning models considered in this work. Because the number of
training samples (labeled data) is limited for many geoscience problems
including the one at hand, we tested several techniques to improve
the performance of CNN models, including (a) augmenting the NOAH TWSA
training samples with additional predictors, such as precipitation
(P) and temperature (T), (b) including regions outside the study area
(i.e., spanning 60\textendash 100$\text{\textdegree}$ longitude, 7.75\textendash 47.75$\text{\textdegree}$
latitude, as shown in Figure \ref{fig:1}) in training to reduce potential
boundary effects and increase training information, and (c) transfer
learning, which \textquotedblleft borrows\textquotedblright{} the
weights from a CNN model trained using many other images. Precipitation
and temperature are already part of the NOAH forcing. The logic behind
including them as additional predictors is that not all the information
in the forcing is fully utilized by the LSM. For example, precipitation
contributes to surface water and groundwater recharge that are not
simulated by NOAH. Similarly, temperature is a proxy of evapotranspiration,
which may not be simulated accurately by the model. \citet{Humphrey2017}
suggested that at least 40\% of the total variance of GRACE anomalies
can be reconstructed from precipitation and temperature variability
alone. Thus, in this study precipitation and temperature are explored
as additional predictors to help improve the model prediction. 
\begin{center}
\begin{figure}[h]
\centering{}\includegraphics[width=4in]{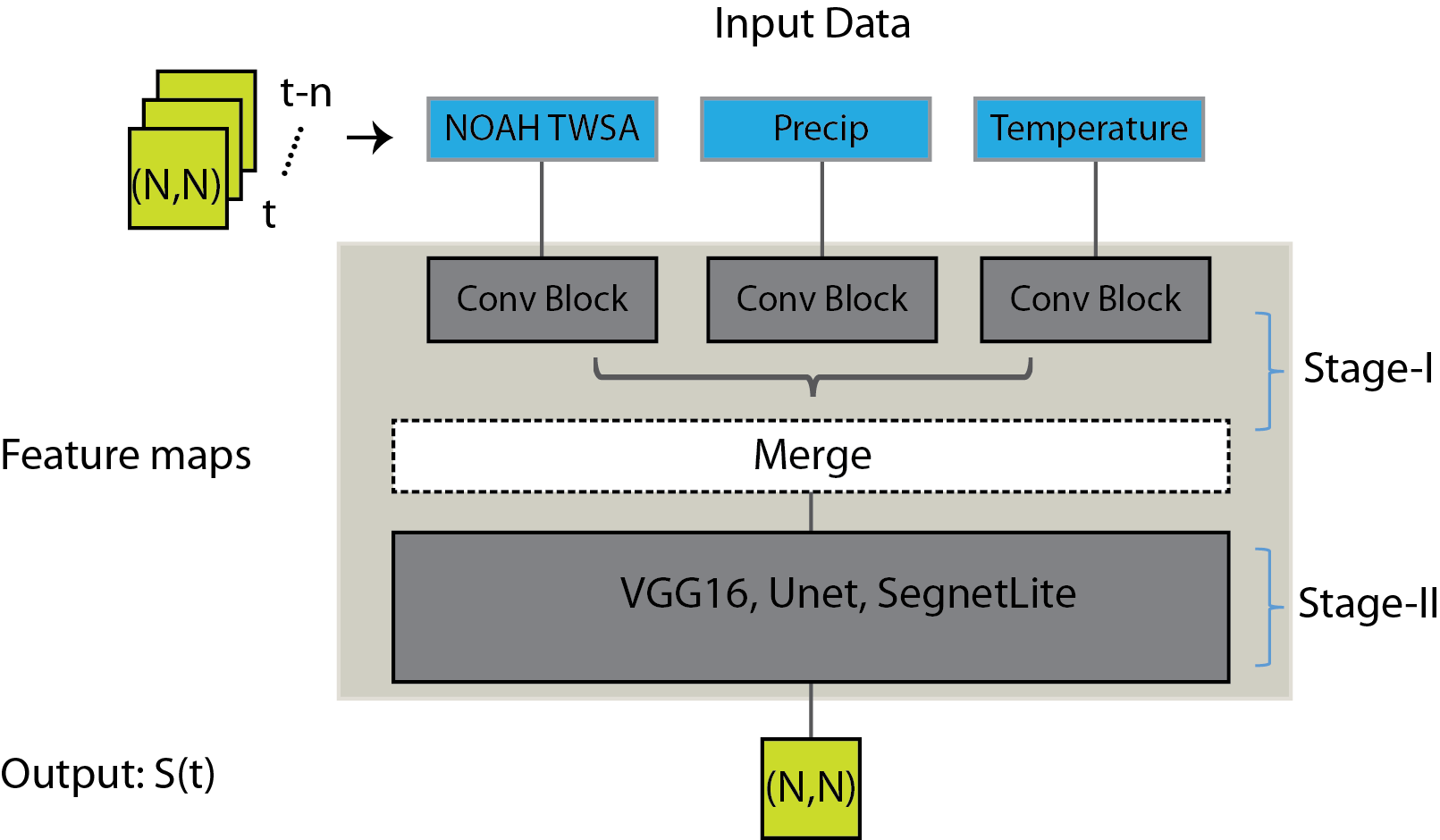}\caption{General CNN model architecture used in this study. The input layer
consists of the NOAH TWSA as the base input stack. Auxiliary predictors
include precipitation and temperature. Each stack of input images
include data from multiple time steps, $t,t-1,\dots,t-n$. The operations
include two stages for shallow and deep learning. The output is the
predicted $S(t)$ having the same dimensions as the input.\label{fig:3}}
\end{figure}
\par\end{center}

As part of data preparation, all input data are formatted or resampled
into 2D images of equal dimensions. Specifically, the 40$\text{\textdegree}$\texttimes 40$\text{\textdegree}$
square region used in this study is represented by 128\texttimes 128
pixel images (0.3125$\text{\textdegree}$ per pixel). The input and target images
are normalized before training. Hydroclimatic variables typically
exhibit certain temporal correlation. To enable the CNN to explore
temporal correlation between each input variable and its antecedent
conditions, we stack the input image at time $t$ on top of its antecedent
conditions to form a 3D volume (see Figure \ref{fig:3}). We set the
number of lags to 2 (i.e., $t-1,\:t-2$) after preliminary experiments;
thus each input volume has dimensions 128\texttimes 128\texttimes 3.
Figure \ref{fig:3} shows that our model design includes two learning
stages. In Stage I, each input volume goes through a separate stack
of convolutional layers. In Stage II, feature maps resulting from
Stage I are merged and the results are fed to a deep learning model
to arrive at the final outputs. The first stage aims to extract unique
features from each input, while the second stage aims to perform deep
learning of the spatial and temporal patterns within each input, as
well as co-variation patterns across the inputs. Putting in a different
way, the role of Stage I is to prepare inputs for use with the problem-independent,
established CNN model architectures employed in Stage II. 

In this work, we consider three CNN-based model architectures, VGG16,
Unet, and Segnet, commonly used in image semantic segmentation problems
(i.e., associating each pixel of an image with a class label). VGG16
is a CNN-based model architecture consisting of 16 layers of 3\texttimes 3
convolutional layers, 2\texttimes 2 max pooling layers, and then a
fully connected layer at the end (Appendix A1). The number of filters
used in each VGG16 convolutional layers monotonically increases. A
VGG16 model pre-trained using 1.3 million images from the ILSVRC-2012
dataset \citep{Simonyan2014} is adopted in this work to implement
transfer learning. In $\mathtt{Keras}$, this is equivalent to freezing
all the hidden layers in VGG16, except for the last fully connected
layer, during training. This way, the CNN model will be able to adjust
itself to the user-specific inputs while transferring most of the
weights learned from the ILSVRC-2012, which includes labeled images
of 1000 object classes \citep{Russakovsky2015}. Previously, \citet{Jean2016}
used transfer learning models to predict poverty based on satellite
imagery. They showed that transfer learning \textquotedblleft can
be productively employed even when data on key outcomes of interest
are scarce.\textquotedblright{} Questions remain about the general
applicability of transfer learning to satellite images, which are
very different from the images used in the ILSVRC dataset. 

Unet has demonstrated superb performance on semantic segmentation
problems, especially on relatively small training datasets \citep{Ronneberger}.
Unet belongs to a class of encoder-decoder model architectures. It
consists of an encoding path (downsampling steps) to capture image
context, followed by a symmetric decoding path (upsampling steps)
to enable precise localization (Appendix A2). The Unet model architecture
used in this study is shown in Appendix A2. It consists of repeated
applications of two 3\texttimes 3 convolution operations, each followed
by a 2\texttimes 2 max pooling layer. The number of filters used is
doubled after each downsampling step and then halved after each upsampling
step. In the final step, a $1\times1$ convolutional layer is used
to generate the output. Unet models are characterized by the copy
and concatenation operations that combine the higher resolution features
from the downsampling path with the upsampled features at the same
level to better localize and learn representations (dashed line with
arrow in Figure \ref{fig:a2}). This is also the part of Unet that
enables multiscale learning. 

Segnet is also a class of encoder-decoder architecture that was originally
introduced to solve image segmentation problems \citep{badrinarayanan2015segnet}.
Similar to the Unet architecture, it includes an encoding path and
a decoding path. The main difference between the design of the original
Segnet and Unet is that the decoder in Segnet uses pooling indices
computed in the max-pooling step of the corresponding encoder to perform
non-linear upsampling, while in Unet the concatenation step is done
before the pooling step. Thus, the number of parameters of Segnet
is smaller than that in the Unet. In this work, we use a variant of
the Segnet architecture, in which the pooling layers are removed and
the upsampling layers in the decoder are replaced by transpose convolution
layers, which may be regarded as performing the reverse of convolutional
operations \citep{zeiler2010deconvolutional}. Different from upsampling,
transpose convolution layers have trainable parameters. The model
design is shown in Appendix A3, which we shall refer to as the SegnetLite
in the rest of this discussion. Similar to Unet, SegnetLite also uses
concatenation steps to combine feature maps from encoding and decoding
steps. The SegnetLite model has a significantly smaller number of
trainable parameters (\textasciitilde 700 thousand) than Unet (7.8
million) and VGG16 (\textasciitilde 134 million ), and can be trained
more efficiently.

\textcolor{black}{For Unet and SegnetLite models, Stage I shallow
learning (Figure \ref{fig:3}) includes a single convolutional layer
with 16 filters for each type of predictors, the outputs of which
are then merged and provided as inputs to the respective deep learning
model. In the case of VGG16, the maximum number of filters that can
be used in Stage I is 3. This is because the trained VGG16 is designed
to process images, which only allow 3 color channels (RGB).}

\subsection{Training and testing of CNN models}

The open-source Python package \texttt{Keras} with the \texttt{Tensorflow}
backend \citep{chollet2015} is used to develop all CNN models presented
in this work. Unless otherwise specified, the stochastic gradient
descent optimizer is used to train the CNN models with a learning
rate of 0.01, decay rate of $1\times10^{-6}$, and momentum of 0.9.
Out of a total of 177 monthly data available for the study period,
125 months or 70\% is used for training and the rest for testing.
The loss or objective function used for network training is the weighted
sum of two fitting criteria
\begin{equation}
\begin{array}{cc}
\textrm{Criterion 1:} & \frac{1}{N_{g}N}\sum_{i=1}^{N}\sum_{j=1}^{N_{g}}(f_{i,j}-S_{i,j})^{2},\\
\\
\textrm{Criterion 2:} & \frac{\sum_{i=1}^{N}\sum_{j=1}^{N_{g}}\left|f_{i,j}-S_{i,j}\right|}{\sum_{i=1}^{N}\sum_{j=1}^{N_{g}}\left|S_{i,j}-\bar{S}_{j}\right|},
\end{array}\label{eq:7}
\end{equation}

\noindent in which $f_{i,j}$ and $S_{i,j}$ are the predicted and
observed mismatch at grid cell $j$ and month $i$, $\bar{S_{j}}$
denotes temporal average at cell $j$, $N_{g}$ is the number of grid
cells in the study area, $N$ is the number of training samples in
the training period, and the summation is taken both spatially and
temporally. Criterion 1 is the commonly used mean square error (MSE)
and Criterion 2 is a modified form of the Nash-Sutcliff efficiency
(NSE) that is more sensitive to over- or underprediction than the
L2 forms used in NSE \citep{krause2005comparison,sun2015model}. The
weight between two criteria is a hyperparameter and is set to 0.5
in this work. 

The performance of trained models is evaluated against the observed
GRACE TWSA using correlation coefficient and NSE. For spatially averaged
time series, the NSE is defined as
\begin{equation}
\text{NSE}=1-\frac{{\displaystyle \sum_{i}^{N_{v}}\left(\text{TWSA}_{\text{GRACE},i}^{\circ}-\left(\text{TWSA}_{\text{NOAH},i}^{\circ}-f_{i}^{\circ}\right)\right)^{2}}}{{\displaystyle \sum_{i}^{N_{v}}\left(\text{TWSA}_{\text{GRACE},i}^{\circ}-\langle\text{TWSA}_{\text{GRACE},i}^{\circ}\rangle\right)^{2}}},
\end{equation}
in which ($\text{\textdegree}$) denotes spatially-averaged quantities and $\langle\rangle$
denotes the temporal mean of observed values, and $N_{v}$ is the
number of samples used for evaluation. The range of NSE is $(-\infty,$1{]}. 

All experiments are carried out on a Linux machine (Dell PowerEdge
R730 server) running with GPU (NVIDIA Tesla K80 GPU, 24Gb RAM total).
Training typically takes 4s, 3s, and <1s per epoch for VGG16, Unet,
and SegnetLite, respectively. Epoch is a deep learning term that refers
to a full pass through a given training dataset and each epoch may
include several iterations as determined by the batch size (i.e.,
the number of samples passed to the neural network during each training
iteration).

\section{Results }

Figure \ref{fig:4} shows the seasonal patterns of $S(t)$, obtained
by averaging the grid values over seasons Dec-Jan-Feb (DJF), Mar-Apr-May
(MAM), Jun-Jul-Aug (JJA), and Sep-Oct-Nov (SON). Recall that $S(t)$
represents the mismatch between NOAH and GRACE TWSA which, according
to its definition in Eq. \ref{eq:2}, tends to be negative in wet
seasons and positive in dry seasons because of the missing SWS and
GWS components in NOAH. Significant spatial and temporal variability
can be observed in Figure \ref{fig:4}. In particular, the histograms
plotted on the right panel of Figure \ref{fig:4} suggest that in
MAM (pre-monsoon dry season) and JJA (first part of monsoon season)
$S(t)$ is dominated by positive values with a mean value of 5.1 cm
and 6.3 cm, respectively. The distribution in MAM is positively skewed,
while it is negatively skewed in JJA, suggesting a transition from
dry to wet season. In SON (late in monsoon season) and DJF (post-monsoon
wet season), the pattern of $S(t)$ is dominated by negative values
with a mean of -6.0 cm and -2.3 cm. The negative values cover most
of the regions in central and southern India. The distribution of
$S(t)$ in SON is also distinctively bimodal.
\begin{center}
\begin{figure}[h]
\begin{centering}
\includegraphics[width=5in]{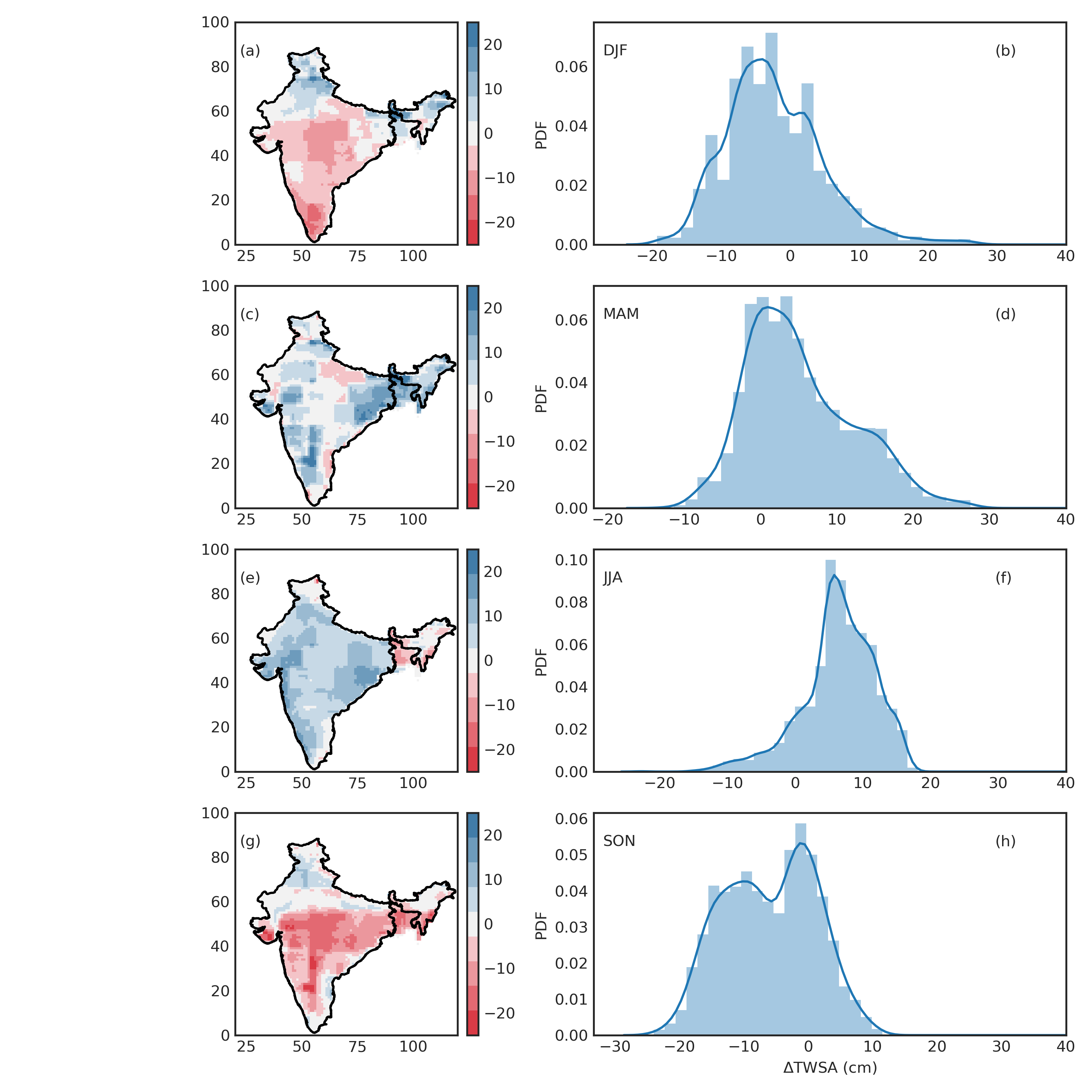}
\par\end{centering}
\caption{Spatial distribution (left panel) and histogram (right panel) of NOAH
and GRACE TWSA mismatch, $S(t),$ averaged over 4 seasons: (a), (b)
DJF; (c), (d) MAM; (e), (f) JJA; and (g), (h) SON. Solid lines on
histograms correspond to fitted PDFs. Map colors are scaled between
(-25cm, 25cm) for visualization.\label{fig:4}}
\end{figure}
\par\end{center}

In the base case, we test the performance of VGG16, Unet, and SegnetLite
models using only NOAH TWSA as the predictor (Table \ref{tab:1}).
The CNN-corrected TWSA is obtained by subtracting the predicted $S(t)$
time series from the NOAH-simulated TWSA using Eq. \ref{eq:2}. For
comparison purposes, all models are trained over 60 epochs with a
batch size of 5. Increasing the number of epochs further did not improve
the results in our experiments. For each of the three CNN models,
the correlation coefficient and NSE between the predicted and GRACE
TWSA at both the country level and grid level are compared. This is
because the GRACE research community is mostly interested on large-scale
averaged results. Note that the actual training is done at the grid
or pixel level, while the country-level statistics are calculated
using grid-averaged TWSA time series. The country-level results are
summarized in Table \ref{tab:1}. For comparison, the metrics between
the original NOAH TWSA and GRACE TWSA are reported in the first row. 

At the country level, all CNN models achieved high correlation (>0.98)
during training, which are all significantly higher than the correlation
between the original NOAH TWSA and GRACE TWSA (0.78). For the testing
period, the correlation values decrease slightly to about 0.94 on
average, but are still higher than the correlation between the original
NOAH and GRACE (0.83), or a 14\% improvement on average. Because NOAH
TWSA, GRACE TWSA, and $S(t)$ are correlated, we applied Williams
significance test \citep{williams1959regression} to test the improvement
in correlation due to deep learning. The p-value of the Williams test
is <0.002 for all three models (see Supporting Information (SI) S1),
suggesting statistically significant improvement. It is worth noting
that the correlation results obtained in this study are comparable
to that obtained by \citet{Girotto2017}, who reported that data assimilation
increased the correlation between their model-simulated TWSA and GRACE
to a country average of 0.96. Correlation coefficient measures the
degree to which model and observations are related in phase, while
NSE, a measure of predictive power, is sensitive to matches (or mismatches)
of both magnitude and phase between the predicted and observed time
series. In this case, the NSE value of the original NOAH TWSA is relatively
low (0.6) for the training period. Figure \ref{fig:5}a plots the
base case results (solid lines in color), the GRACE TWSA (dark solid
line with filled circles) and its error bound (shaded area), and the
uncorrected NOAH TWSA (gray dashed line). For the training period,
the plot suggests that the uncorrected NOAH TWSA underestimates most
of the wet and dry events. In contrast, both Unet (orange line) and
SegnetLite (green line) fit the wet and dry events well and are within
the extent of the GRACE data uncertainty. The VGG16 model (dark blue
solid line) underestimates the magnitudes of some wet events in 2002,
2003, and 2007. 

During the testing period, we see several dry events, for example,
the severe droughts in 2013 and 2016. In the literature, the dry events
in 2014 and 2015 were attributed to monsoon rainfall deficits \citep{mishra2016frequency}.
Again, the uncorrected NOAH underestimates the dry and wet events,
especially the dry events. The SegnetLite model captures all dry events
in 2013\textendash 2016 well, but slightly underestimates the 2014
and 2015 wet peaks. On the other hand, the VGG16 model captures most
of the wet events, but underestimates dry events. The performance
of Unet is in between. The average NSE improvement in the testing
period is 0.87, or 52\% improvement over the uncorrected NOAH TWSA.
Figure \ref{fig:5}a also suggests that even though the CNN models
are trained at the grid level, they conserve mass at the country level.
This is encouraging and may be attributed to the strong ability of
CNN to learn multiscale spatial features and, therefore, preserve
spatial continuity inherent in the input. 

Figures \ref{fig:5}b and \ref{fig:5}c show the cumulative distribution
function (CDF) of the pixel-wise, or grid-scale correlation coefficient
and NSE between modeled TWSA and GRACE TWSA. The CDFs of all CNN-corrected
results (solid lines in color) show a clear improvement over the original
NOAH model (dashed line). Both Unet and SegnetLite give better performance
than VGG16 and, in particular, the performance of SegnetLite is slightly
better in the upper range of the correlation coefficient and NSE CDFs.
The results thus far suggest that the mismatch pattern learned using
NOAH TWSA as the base predictor can already help to correct the NOAH
results significantly, both in magnitude and phase. On the basis of
Table \ref{tab:1} and Figure \ref{fig:5}, the SegnetLite model shows
the best performance for the base case. The VGG16 model gives slightly
worse results than the other two, probably because of the limited
number of input feature maps it allows. 
\begin{center}
\begin{figure}[h]
\centering{}\includegraphics[width=5in]{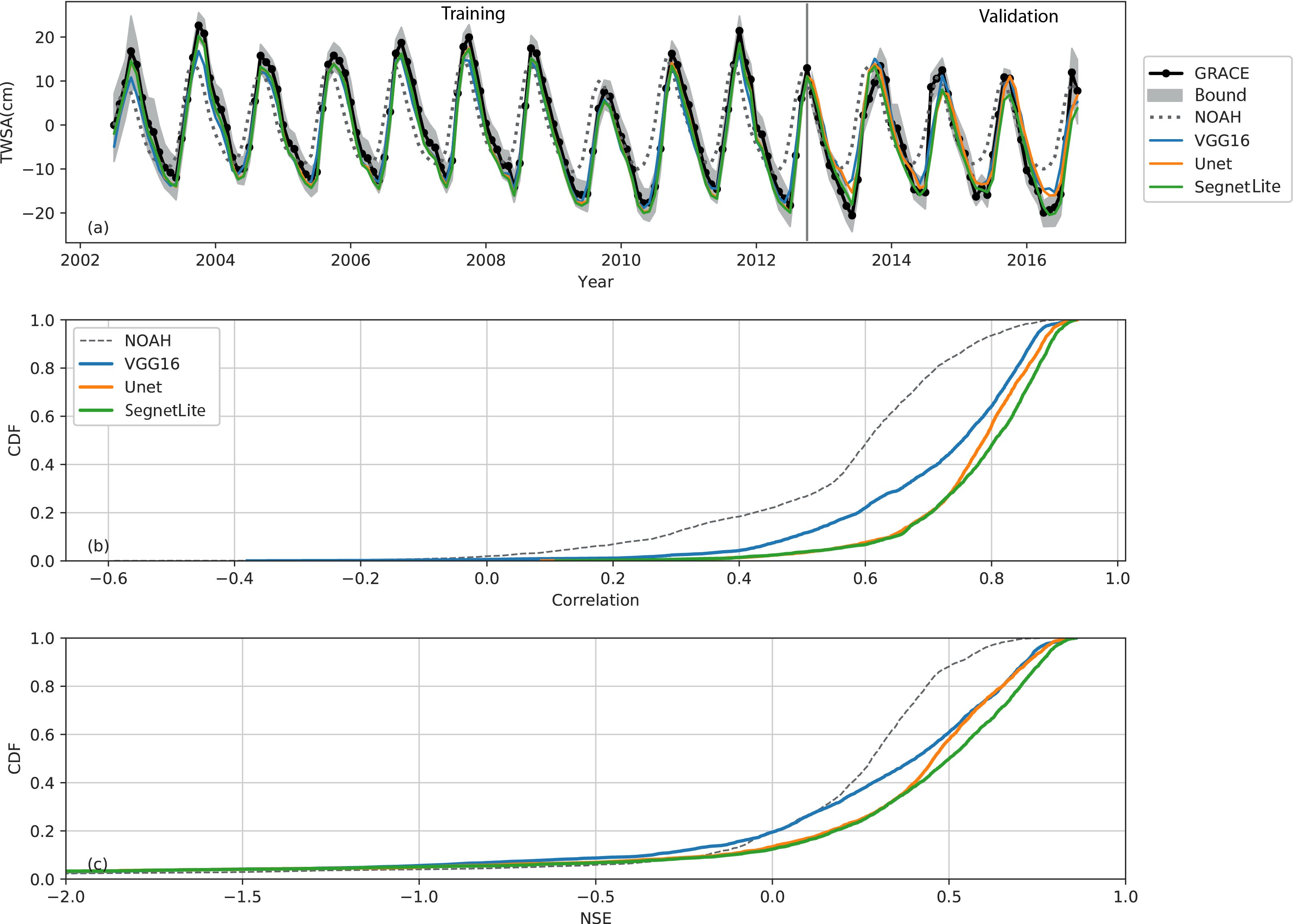}\caption{Comparison of (a) GRACE (dark solid line with filled circles), NOAH
(gray dashed line), and CNN-corrected TWSA time series by VGG16 (blue),
Unet (orange), and SegnetLite (green) for training and testing periods
(separated by the thin vertical bar) at the country level; (b) and
(c) CDFs of correlation coefficient and NSE between modeled TWSA (including
both NOAH and CNN-corrected results) and GRACE at the grid level.
Shaded area in (a) represents the total error bound estimated for
GRACE TWSA (see Section 2). \label{fig:5}}
\end{figure}
\par\end{center}

\begin{figure}[H]
\centering{}\includegraphics[width=5in]{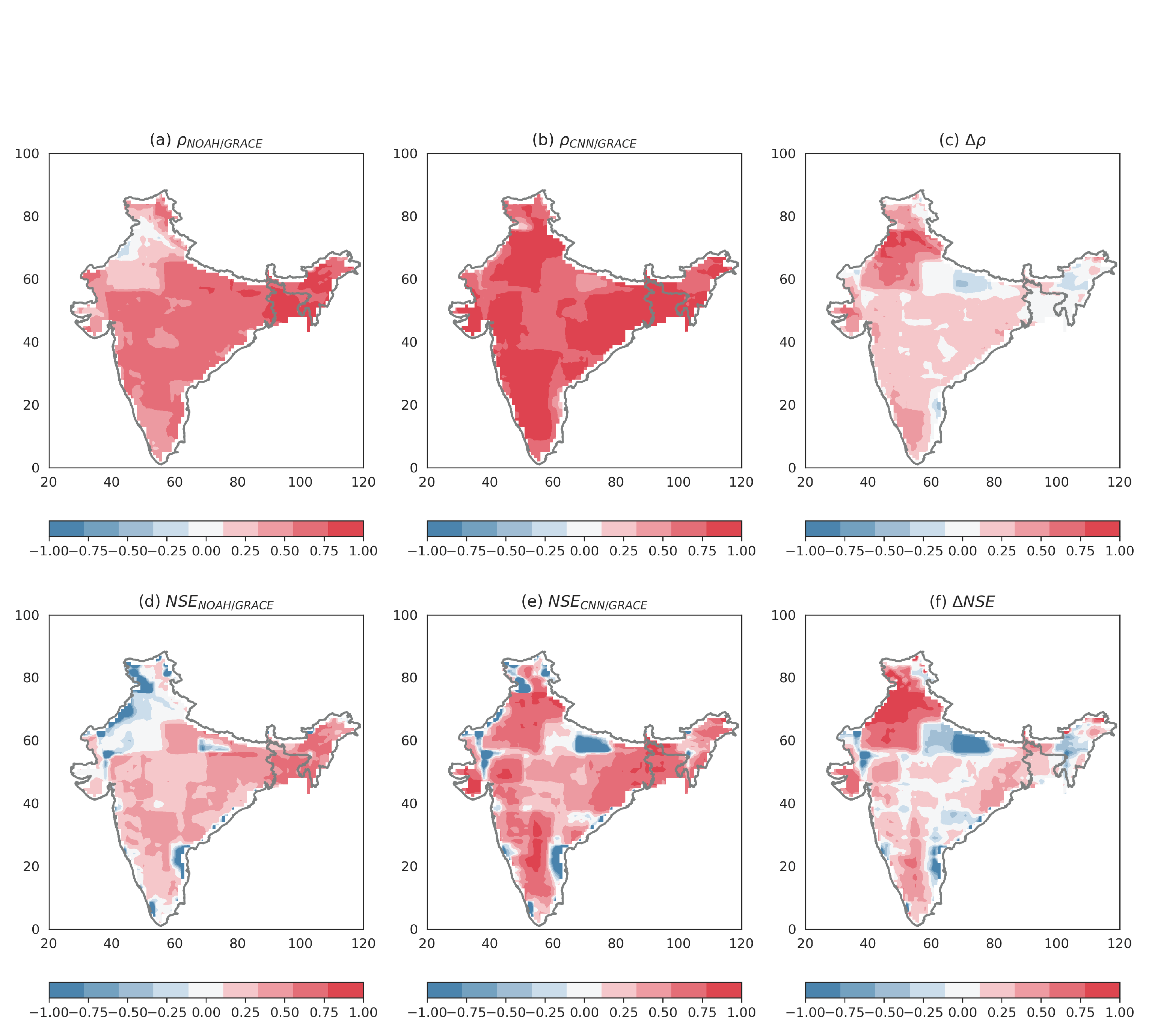}\caption{Grid-scale correlation coefficient maps between (a) NOAH-simulated
TWSA and GRACE, (b) SegnetLite corrected TWSA and GRACE, and (c) the
difference between (a) and (c); (d)\textendash (f) the same maps but
for NSE. For plotting purposes, all maps are scaled to {[}-1,1{]}.
\label{fig:6}}
\end{figure}

To help interpret the learned spatial patterns further, in Figure
\ref{fig:6} we plot correlation and NSE maps corresponding to the
uncorrected NOAH TWSA (\ref{fig:6}a, \ref{fig:6}d), the SegnetLite
model (\ref{fig:6}b, \ref{fig:6}e), and improvements due to CNN
correction, for the period 2002/04\textendash 2016/12 (\ref{fig:6}c,
\ref{fig:6}f). In general, higher correlation and NSE values are
observed in southcentral and central India. The correlation improvement
is the greatest in northwest and south India. The drier northwest
India has been significantly affected by anthropogenic activities
related to irrigation, whereas the wetter southmost part of the country
is subject to bimodal precipitation pattern \citep{Girotto2017},
both are not resolved well in the current NOAH model. On the other
hand, regional groundwater impact related to water withdrawal in northwest
India has been confirmed by a number of previous GRACE studies \citep[e.g., ][]{Rodell2009,Chen2014}.
Thus, the TWSA correction benefits the most in those areas. Nevertheless,
isolated weak spots, especially on NSE maps, are found near the India-Nepal
border (part of Ganges River Basin) and also in the Brahmaputra River
Basin, where NOAH already gives good performance and the improvements
by CNN are either insignificant or even deteriorated. The Himalayas
region outside India's north border may have negative impact on the
learning because of sharp discontinuity in patterns. Similarly, the
isolated weak spots along the Indian coast may also be related to
the lack of continuity in patterns. Additional data may be necessary
to constrain the CNN learning in those isolated spots. To give a sense
of fitting quality, we show grid-level time series of NOAH TWSA, GRACE
TWSA, and SegnetLite corrected TWSA at four selected pixel locations
in SI S2. Two examples correspond to locations of significant NSE
improvement (northwest India and southcentral India) and the other
two examples show locations of performance deterioration (India-Nepal
border and southern coastal area). SI S2 suggests that at the northwest
India location, deep learning helps to improve the match of a downward
trend observed by GRACE. SI S3 plots the same maps as shown in Figure
\ref{fig:6} but for the testing period 2012/09\textendash 2016/12.
In general, the same improvement patterns (i.e., $\Delta\rho$ and
$\Delta\mathrm{NSE}$) are observed over most of the region, except
for north India where the effect due to correlation correction is
little or none. The absolute NSE over northwest India is lower than
that in Figure \ref{fig:6}, although the NSE correction is still
significant over most of the study region.

We performed additional tests for each type of CNN models by adding
precipitation (P) and temperature (T) as predictors (Table \ref{tab:1}).
Results show that the additional predictors have little improvement
over the base case (SI S4). Although P and T may include additional
information (e.g., on SWS) not already included in the model, their
effect may be limited by the resolutions of CNN models and GRACE observations,
and by the strong seasonality of the study area. Nevertheless, P and
T forcing may still be useful for reconstructing the TWS for other
parts of the world.

To further corroborate the learned patterns, we now compare $S(t)$
to in situ groundwater storage anomalies (GWSA). As mentioned before,
NOAH does not include SWS and GWS, while GRACE observes the total
water column in space. Thus, the mismatch pattern should reflect the
missing components, and is expected to correlate well with in situ
GWSA wherever the TWSA is dominated by GWS. We assign groundwater
wells to the nearest CNN model grid cells and then calculate the correlation
coefficient between $S(t)$ estimated by SegnetLite and in situ GWSA.
Results are shown in Figure \ref{fig:7}. Spatially, positive correlations
are observed for most parts of India. The 50th percentile of correlation
is about 0.4 (inset of Figure \ref{fig:7}). The correlation is weaker
in northwest India, the India-Nepal border, and along the southern
coastal areas. The weaker correlation in northwest India is intriguing,
given the dominance of groundwater in that region and strong correlation
between the corrected NOAH and GRACE TWSA obtained for the same area,
as shown in Figure \ref{fig:6}b. One possible explanation is given
by \citet{Girotto2017}, who pointed out that groundwater used for
irrigation in northwest India is \textquotedblleft extracted primarily
from deep aquifers, which are observed by GRACE, but not by the shallow
in situ groundwater measurements.'' Thus, the limitation of the in
situ dataset needs to be kept in mind when interpreting the comparison
results in Figure \ref{fig:7}. For areas along the Indus River and
Ganges River, the impact of surface water is relatively high \citep{getirana2017rivers},
which limits the proportion of GWSA in $S(t)$ and weakens the correlation
between $S(t)$ and in-situ GWSA. Note in this comparison with the
in situ GWSA, we mainly focus on analyzing the phase agreement because
of the uncertainty of in situ GWSA magnitudes related to the uncertain
specific yield.
\begin{center}
\begin{figure}[H]
\centering{}\includegraphics[height=5in]{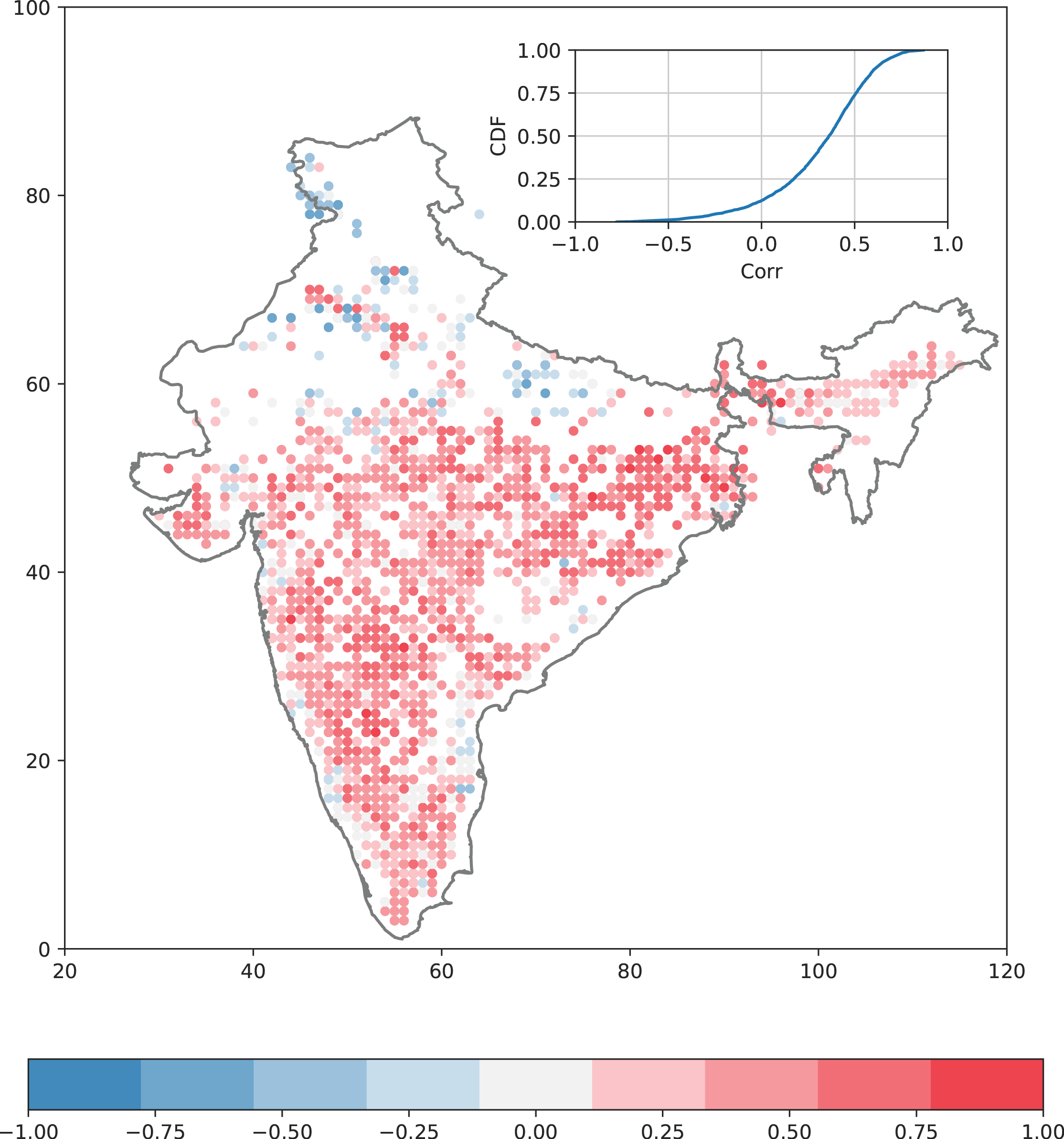}\caption{Correlation map between in situ GWSA and $S(t)$ learned by SegnetLite
model. Inset shows the CDF of correlation coefficient. The map coordinates
are grid cell indices (from 0 to 127). \label{fig:7}}
\end{figure}
\par\end{center}

Finally, we apply the trained SegnetLite model to predict TWSA. Figure
\ref{fig:8} shows the country-averaged TWSA for the period 2016\textendash 2017.
The GRACE data (green filled circles) becomes unavailable after June
2017. Also, the GRACE data from 2017 is not part of the model training
and testing. The 95\% prediction interval is estimated using 1.96
RMSE, where the RMSE (\textasciitilde{} 2.20 cm) is calculated by
using the misfit of SegnetLite model on the training data. The plot
suggests that the SegnetLite model captures the GRACE data well during
the 2017 months that are not part of Figure \ref{fig:5}a, demonstrating
the potential use of this method for filling data gaps between GRACE
and GRACE-FO.
\begin{center}
\begin{figure}[H]
\centering{}\includegraphics[width=4in]{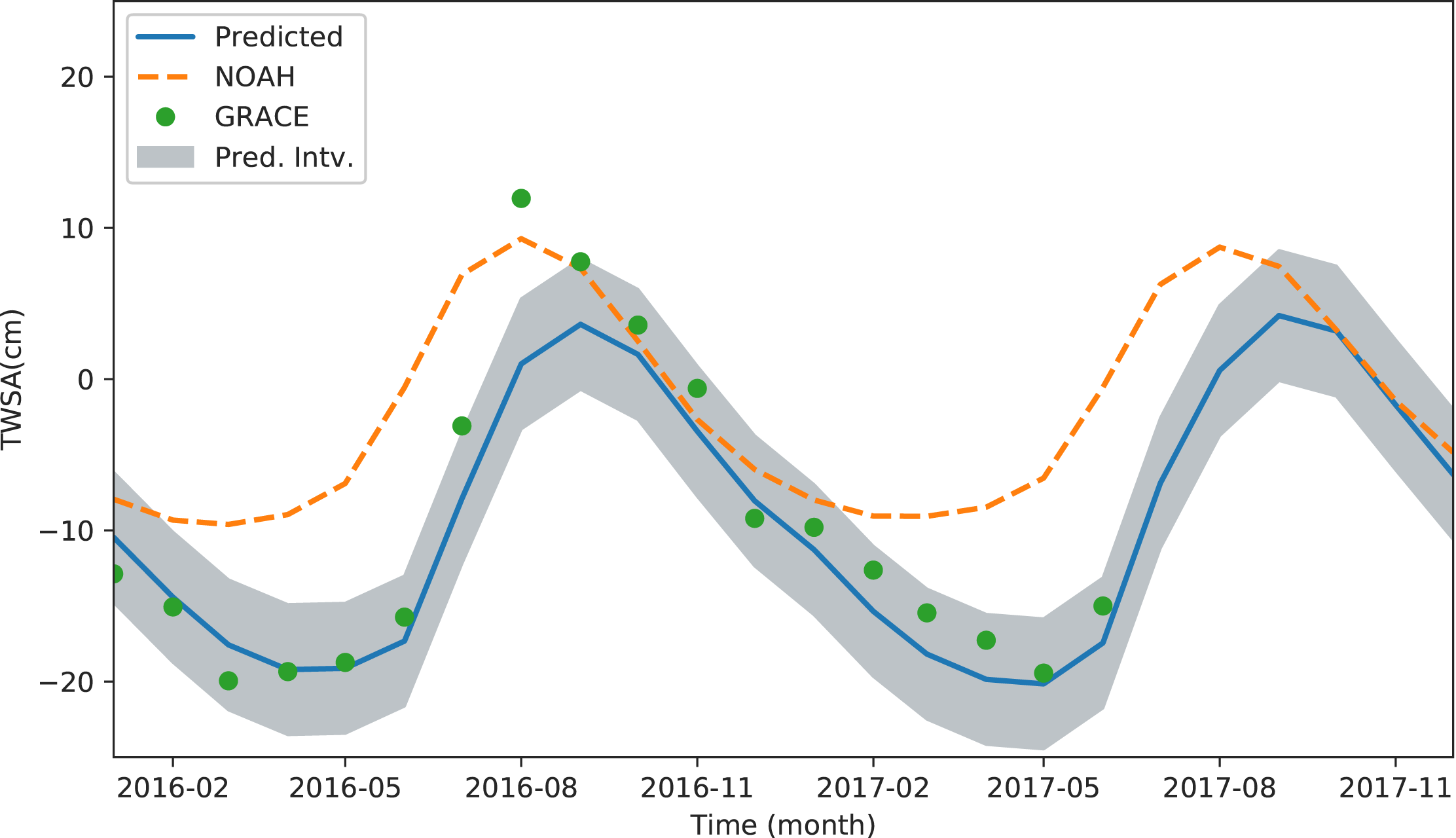}\caption{Country averaged TWSA (blue solid line) predicted for 2016\textendash 2017
by using the trained base SegnetLite model. Dashed line (orange) is
the NOAH TWSA output, and also the input to the SegnetLite model.
Filled circles (green) represent GRACE monthly data, and shaded area
corresponds to 95\% prediction intervals. The vertical line marks
the beginning of ``unseen'' data during previous training and testing.
\label{fig:8}}
\end{figure}
\par\end{center}

\section{Conclusion}

In this study, we present a hybrid approach that combines physically-based
modeling and deep learning to predict the spatial and temporal variations
of TWS anomaly (TWSA). This is done by training CNN-based deep learning
models (VGG16, Unet, and SegnetLite) to learn the spatial and temporal
mismatch pattern between the TWSA simulated by a land surface model,
NOAH, and that observed by GRACE, using which the NOAH-simulated TWSA
is then corrected. The hybrid modeling approach is systematically
demonstrated over India by using various performance metrics. In general,
all deep learning models considered in this study are able to improve
the NOAH TWSA significantly at both the country- and grid level, which
is encouraging because we deal with a much smaller training sample
size than those typically used in image classification problems. A
correlation analysis between the learned patterns and the in situ
groundwater storage anomaly (GWSA) shows good correlation between
the two, suggesting the learned patterns effectively compensate for
the missing groundwater storage in NOAH for many parts of the study
area. 

Our method presents an alternative for extrapolating TWSA time series
outside the GRACE period. Our results also indicate the feasibility
of using deep learning to perform spatial and temporal interpolation,
which has long been a challenging problem in the geoscience literature.
Compared to the conventional 4D variational or ensemble-based data
assimilation techniques for fusing hydroclimatic data, major strengths
of our hybrid approach include (1) the relatively few assumptions
involved, especially with regard to parameterization of the spatial
and temporal error distributions; (2) the capability to extract useful
features at multiple scales, and (3) the capability to handle multiple
data types with relative ease. 

Deep learning algorithms evolve rapidly. In this study, we mainly
considered three variants of CNN. In the literature, long short-term
memory (LSTM) and recurrent neural networks (RNN) have been combined
with CNN for spatiotemporal prediction problems \citep{shi2015convolutional,Fang2017}.
In addition, the grid resolution of our networks is relatively coarse.
Finer resolution grids may be tested in the future to improve model
fits. 

\section*{Appendix}

\setcounter{figure}{0} \renewcommand{\thefigure}{A.\arabic{figure}}

\subsection*{A.1 VGG-16}

\begin{figure}[H]
\centering{}\includegraphics[width=5in]{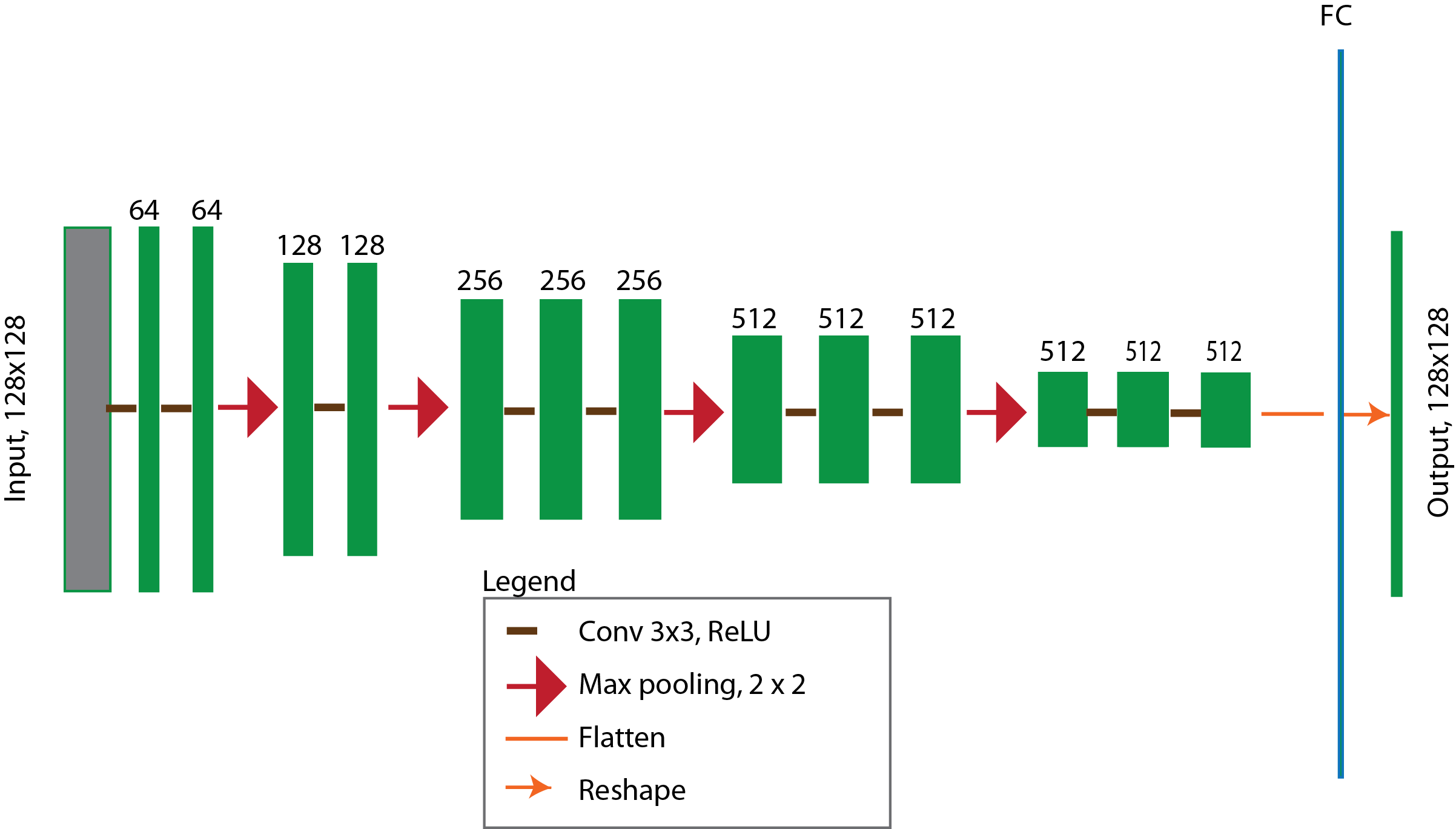}\caption{VGG16 model architecture. \label{fig:a1}}
\end{figure}

The pre-trained VGG16 model (i.e., weights) is obtained from the \texttt{Keras}
package \citep{chollet2015}. The VGG16 model design consists of a
series of downsampling convolutional layers (3\texttimes 3 filter,
ReLU activation function), interlaced with max pooling layers (2\texttimes 2)
(Figure \ref{fig:a1}). The number of filters increases gradually
from 64 to 512, while the size of the feature map decreases from 128
to 8 (in pixels). At the end, the convolutional layers are flattened
and connected to a fully connected layer before reshaped to the dimensions
of the output layer (i.e., 128\texttimes 128). Linear activation function
is used for the output layer.

\subsection*{A.2 Unet model}

\begin{figure}[H]
\centering{}\includegraphics[width=5in]{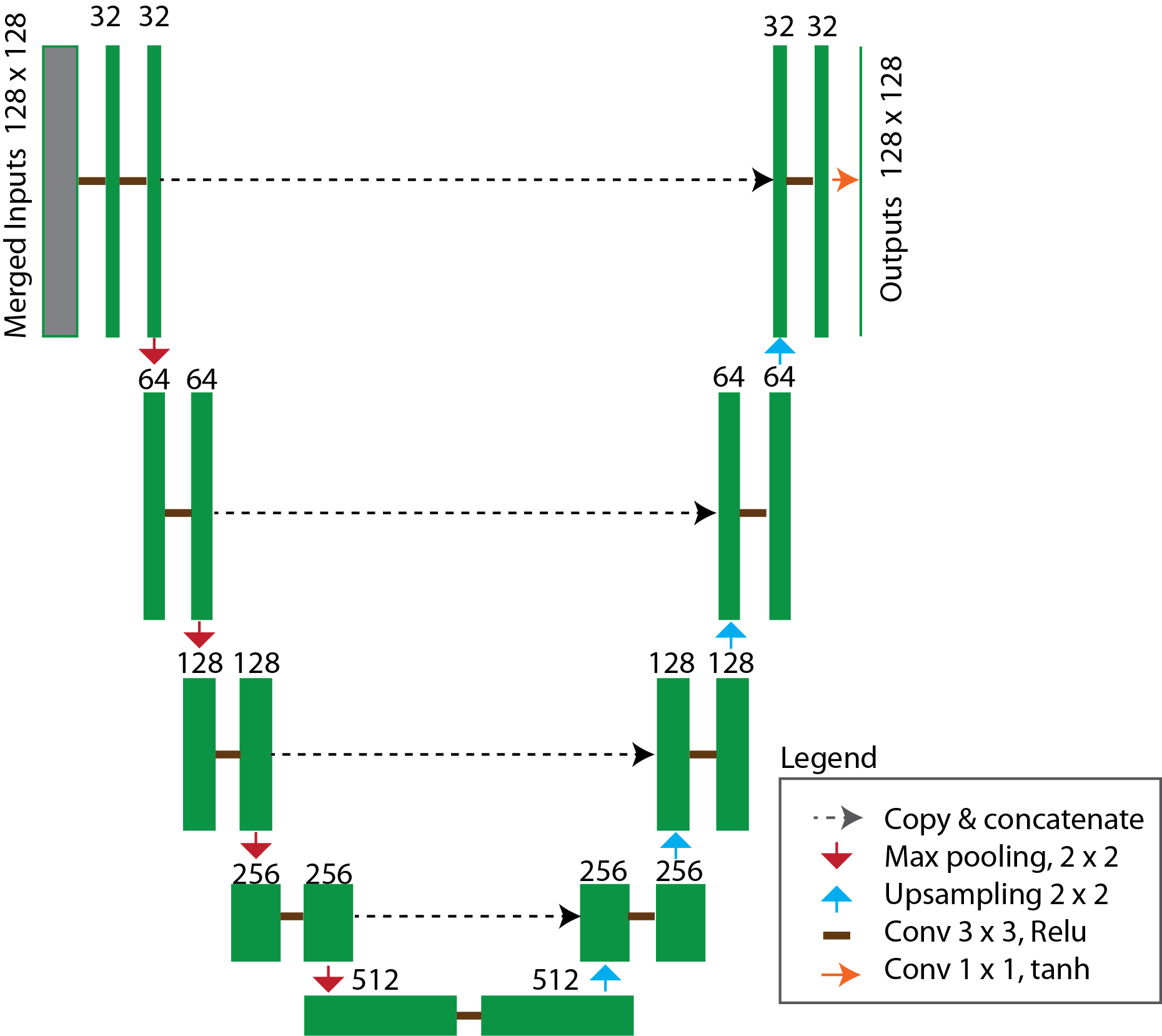}\caption{Unet model architecture. \label{fig:a2}}
\end{figure}

The Unet model used in this work is adapted from the original design
of \citet{Ronneberger}, with modifications in the number of filters
used. The model design belongs to a class of encoder-decoder architectures.
The encoder part includes 5 consecutive downsampling steps, and the
decoder part includes an equal number of upsampling steps. Each downsampling
step involves 2 convolutional layers (using 3\texttimes 3 filter and
ReLU activation function), followed by a max pooling layer (2\texttimes 2).
For the encoder part, the number of filters in the convolutional layers
increases from 32 to 512, while the dimensions of the feature maps
decrease from 128\texttimes 128 to 8\texttimes 8. Each upsampling
step involves (a) an upsampling step (2\texttimes 2), (b) a concatenation
step in which the feature maps from the same level of downsampling
and upsampling paths are combined (dashed line in Figure \ref{fig:a2}),
and (c) two convolutional layers (using 3\texttimes 3 filter and ReLU
activation function). In its simplest form, upsampling repeats rows
and columns to create a larger image (no trainable parameters). The
inputs to the model include image stacks corresponding to one or more
predictors, and the output of the model is the predicted mismatch,
$S(t)$. A $1\times1$ convolutional layer with linear activation
function is used to generate the output.

\subsection*{A.3 SegnetLite model}

\begin{figure}[H]
\centering{}\includegraphics[width=5in]{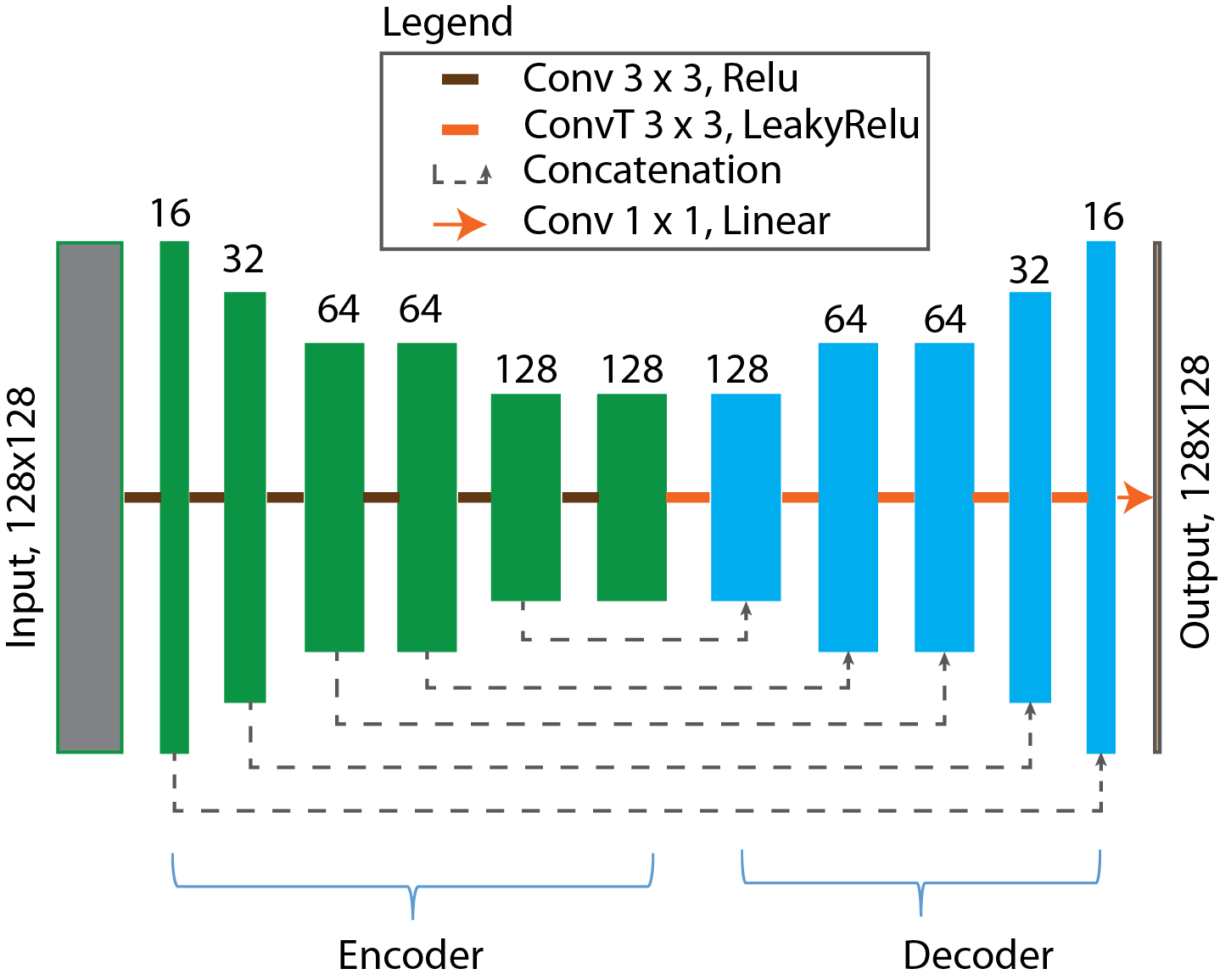}\caption{SegnetLite model architecture.\label{fig:a3}}
\end{figure}

Segnet is a deep CNN architecture introduced to perform semantic segmentation
\citep{badrinarayanan2015segnet}. SegnetLite used in this work is
a variant of the original Segnet, which uses a smaller number of encoding
and decoding steps. In addition, no max-pooling is used and the upsampling
layers in the original design are replaced by transpose convolution
layers, which can be regarded as the reverse of convolutional operations
and which increase the input dimensions like the upsampling does.
However, transpose convolution layers introduce trainable parameters
to learn the optimal upsampling parameters. The encoder part of SegnetLite
consists of six convolution layers, with the number of filters increasing
from 16 to 128, while the decoding part is symmetric and includes
alternating concatenation and transpose convolution layers (Figure
\ref{fig:a3}). Similar to Unet, SegnetLite uses concatenation steps
to combine feature maps from encoding and decoding steps at the same
level. To generate the output layer, an upsampling layer is used to
increase the decoder outputs to the output dimensions ($128\times128)$
and is then passed through a $1\times1$ convolutional layer as in
the other two models.

\section*{Acknowledgments}

The GRACE mascon product used in this study was downloaded from JPL
(\url{www.grace.jpl.nasa.gov}). NOAH data were downloaded from NASA
Earthdata site (\url{http://earthdata.nasa.gov}). Processing and
numerical experiments were carried out on the Chameleon Cloud hosted
by Texas Advanced Computing Center. A. Y. Sun and B. R. Scanlon were
partially supported by funding from Jackson School of Geosciences,
UT Austin. The authors are grateful to the Associate Editor and four
anonymous reviewers for their constructive comments.

\bibliographystyle{agufull08}

\newpage{}
\begin{center}
\begin{table}
\caption{Model performance metrics.\label{tab:1}}

\centering{}%
\begin{tabular}{|c|c|c|c|c|}
\hline 
\multirow{3}{*}{Model} & \multicolumn{4}{c|}{Performance Metrics}\tabularnewline
\cline{2-5} 
 & \multicolumn{2}{c|}{Training} & \multicolumn{2}{c|}{Testing}\tabularnewline
\cline{2-5} 
 & Corr & NSE & Corr & NSE\tabularnewline
\hline 
\hline 
$\text{TWSA}_{\text{NOAH}}$ & 0.776 & 0.600 & 0.825 & 0.568\tabularnewline
\hline 
\multicolumn{5}{|c|}{Base Case, $\text{TWSA}_{\text{NOAH}}$ only}\tabularnewline
\hline 
VGG16 & 0.986 & 0.925 & 0.944 & 0.862\tabularnewline
\hline 
Unet & 1.0 & 0.948 & 0.938 & 0.868\tabularnewline
\hline 
Segnet & 1.0 & 0.952 & 0.946 & 0.875\tabularnewline
\hline 
\multicolumn{5}{|c|}{$\text{TWSA}_{\text{NOAH}}$ and P}\tabularnewline
\hline 
VGG16-2 & 0.986 & 0.909 & 0.928 & 0.861\tabularnewline
\hline 
Unet-2 & 1.0 & 0.969 & 0.941 & 0.880\tabularnewline
\hline 
Segnet-2 & 1.0 & 0.961 & 0.943 & 0.880\tabularnewline
\hline 
\multicolumn{5}{|c|}{$\text{TWSA}_{\text{NOAH}}$, P, and T}\tabularnewline
\hline 
VGG16-3 & 0.985 & 0.906 & 0.936 & 0.864\tabularnewline
\hline 
Unet-3 & 1.0 & 0.961 & 0.939 & 0.876\tabularnewline
\hline 
Segnet-3 & 1.0 & 0.977 & 0.945 & 0.889\tabularnewline
\hline 
\end{tabular}
\end{table}
\newpage{}
\par\end{center}

\begin{flushleft}
\textbf{Figure Captions}
\par\end{flushleft}

\begin{flushleft}
Figure 1. Map of study area (latitude: 7.75\textendash 47.75$\text{\textdegree}$,
longitude: 60\textendash 100$\text{\textdegree}$), where India is bounded by
the dark solid line. During training, data corresponding to the entire
square area is used to reduce potential boundary effects and increase
information content for training.
\par\end{flushleft}

\begin{flushleft}
Figure 2. Illustration of the flow of information from GLDAS-NOAH
and GRACE to the deep learning model. Here the observed mismatch $S(t)$
is only used to train the CNN deep learning model and is no longer
required after the model is trained. NOAH TWSA is the base predictor.
Other predictors may include precipitation and temperature. The dashed
arrow indicates that the same $S(t)$ is also used for GRACE data
assimilation studies. 
\par\end{flushleft}

\begin{flushleft}
Figure 3. General CNN model architecture used in this study. The input
layer consists of the NOAH TWSA as the base input stack. Auxiliary
predictors include precipitation and temperature. Each stack of input
images include data from multiple time steps, $t,t-1,\dots,t-n$.
The operations include two stages for shallow and deep learning. The
output is the predicted $S(t)$ having the same dimensions as the
input.
\par\end{flushleft}

\begin{flushleft}
Figure 4. Spatial distribution (left panel) and histogram (right panel)
of NOAH-GRACE mismatch, $S(t),$ averaged over 4 seasons: (a), (b)
DJF; (c), (d) MAM; (e), (f) JJA; and (g), (h) SON. Solid lines on
histograms correspond to fitted PDFs. Map colors are scaled between
(-25, 25) cm for visualization.
\par\end{flushleft}

\begin{flushleft}
Figure 5.Comparison of (a) GRACE (dark solid line with filled circles),
NOAH (gray dashed line), and CNN-corrected TWSA time series by VGG16
(blue), Unet (orange), and SegnetLite (green) for training and testing
periods (separated by the thin vertical bar) at the country level;
(b) and (c) CDFs of correlation coefficient and NSE between modeled
TWSA (including both NOAH and CNN-corrected results) and GRACE at
the grid level. Shaded area in (a) represents the total error bound
of GRACE TWSA. 
\par\end{flushleft}

\begin{flushleft}
Figure 6. Grid-scale correlation coefficient maps between (a) NOAH-simulated
TWSA and GRACE, (b) SegnetLite-corrected TWSA and GRACE, and (c) their
differences; (d)\textemdash (f) the same maps but for NSE. For plotting
purposes, all maps are scaled to {[}-1,1{]}.
\par\end{flushleft}

\begin{flushleft}
Figure 7. Correlation map between in situ GWSA and $S(t)$ learned
by SegnetLite model. Inset shows the CDF of correlation coefficient.
The map coordinates are grid cell indices (from 0 to 127). 
\par\end{flushleft}

\begin{flushleft}
Figure 8. Country averaged TWSA (blue solid line) predicted for 2016\textendash 2017
by using the trained base SegnetLite model. Dashed line (orange) is
the NOAH TWSA output, and also the input to the SegnetLite model.
Filled circles (green) represent GRACE monthly data, and shaded area
corresponds to 95\% prediction intervals. 
\par\end{flushleft}

\begin{flushleft}
Figure A1. VGG16 model architecture.
\par\end{flushleft}

\begin{flushleft}
Figure A2. Unet model architecture. 
\par\end{flushleft}

\begin{flushleft}
Figure A3. SegnetLite model architecture.\newpage{}
\par\end{flushleft}
\end{document}